\documentclass[12pt]{book}

\usepackage[dvips]{graphicx,color}
\usepackage{makeidx,tsukuba}

\makeauthorindex
\makeindex

\begin{document}

\BookTitle{\itshape The 28th International Cosmic Ray Conference}
\CopyRight{\copyright 2003 by Universal Academy Press, Inc.}
\pagenumbering{arabic}

%
%
%
%

\chapter{The Local Interstellar Spectrum of Cosmic Ray Electrons} 

\author{D.~Casadei and V.~Bindi\\
{\it
 Bologna University and INFN Bologna, via Irnerio 46, I-40126
  Bologna, Italy}
}

\section*{Abstract}

 The local interstellar spectrum of cosmic ray electrons and positrons
 from 0.8 GeV to 2 TeV is derived by demodulating the measured spectra
 by balloon and satellite experiments.  It can be well represented by
 a single power law in kinetic energy with spectral index 3.4 over the
 whole energy range, pushing for the idea that it is not
 representative of the galactic average.  Instead, the spectrum has to
 reflect the nature of our local bubble, being mostly sensitive to the
 last nearby supernova.

\section{Acceleration and propagation of cosmic ray electrons}

 The most favored sites for cosmic ray (CR) acceleration are the shock
 fronts generated by supernova (SN) explosions: the SN emitted power
 and their estimated rate are compatible with the CR energy density in
 the Galaxy (of the order of 1 eV cm$^{-3}$) if the acceleration
 mechanism has an efficiency of few percent, while the Fermi model of
 the acceleration by magnetic irregularities naturally produces a
 power-law spectrum in momentum.  Observation of synchrotron X-rays
 from supernova remnants (SNR) demonstrates that electrons are
 accelerated by SNR [1].  Howewer, there is no definitive proof of
 hadronic acceleration by SNR.

 The upper limit $E_{\mathrm{max}}$ for the energy of particles with
 charge $Ze$ that are accelerated by a SN shock front is found
 considering the finite life time and size of the engines:
 $E_{\mathrm{max}} \sim Z \times 10^{15}$ eV.  The maximum energies
 (about 10 TeV) of detected gamma rays from SNR, interpreted as the
 result of inverse Compton scattering of high energy electrons with
 ambient photons, imply an upper limit to the acceleration of
 electrons well below this theoretical estimate [2].  On the other
 hand, the $\gamma$-rays emitted by SNR could be produced as well by
 the hadronic component (via $\pi^0$ decay): in this case we would
 expect not to observe any cut-off in the $\gamma$-ray spectrum up to
 the knee.  Recently, Guetta \& Amato [3] pointed out that the
 electronic origin of high energy $\gamma$-rays is very probable for
 the Crab and MSH15-52 SNR, whereas the hadronic component is a
 possible explanation of the emission from Vela and G343.1-2.3.

 Electrons are light particles and suffer large energy losses produced
 by electromagnetic processes like bremsstrahlung, synchrotron
 radiation and inverse Compton scattering, whereas all other CR
 particles are relatively insensitive to these processes.  Thus, while
 CR protons and stable nuclei propagate for a large fraction of the
 Galaxy with small energy losses, electrons of energy $E$ diffuse only
 for distances $\sim$ 1 kpc ($E$/TeV)$^{-1}$ [4].  On the other hand,
 the e.m.~radiation can be used to map their distribution (averaged
 along the line of sight) in the Galaxy, suggesting an average source
 spectral index equal to 2 [1].

 Below few GeV, the bremsstrahlung dominates the electron energy
 losses, producing $\gamma$-rays with typical energies of 1--100 MeV.
 If the source injection spectrum is $\propto p^{-\alpha}$, the
 observed spectrum should have a spectral index $\gamma \approx
 \alpha$ at these energies.  Howewer, the electron bremsstrahlung is
 overwhelmed by the $\pi^0$ decay spectrum, due to the hadronic CR
 component [1].  Above 10 GeV inverse Compton and synchrotron
 radiation are the dominant energy losses, producing photons in the X
 and radio bands, respectively.  The observed spectral index is
 expected to be $\gamma = \alpha + 1 - \Delta$, where $\Delta$ is a
 small correction that depends on the source spatial distribution [5].
 Very high energy electrons at the sources can produce synchrotron
 radiation in the X band, thanks to the relatively high magnetic field
 of SNR.  This radiation has a power-law spectrum, with index
 $(\gamma-1)/2$, that stems out of the thermal component, and it is
 localized into well defined structures, like the SN shock fronts [6].

 Because of the strong electromagnetic losses, electrons diffuse away
 from the source for a distance that is inversely proportional to
 their energy.  Hence one may describe their distribution in the
 galactic plane as an ensamble of statistically independent
 populations, whose dimensions are determined by two effects: (1)
 diffusion with average propagation length $\ell (E) = 2 \sqrt{D(E)
 t}$ and spatial diffusion coefficient $D(E) \approx D_0 (1 +
 E/E_*)^\delta$, with $D(10\,\mathrm{GeV}) = $ (1--10)$\times 10^{28}$
 cm$^2$ s$^{-1}$ [5] and $\delta =$ 0.3--0.8 [7], and (2) radiative
 energy losses with characteristic time $t_{\mathrm{rad}} \approx 2.1
 \times 10^5 (E/\mathrm{TeV})^{-1}$ y [4].

 From the observer point of view, the measured CR electron spectrum
 can not be considered representative of any galactic average: the
 spectrum is dominated by the most recent SN explosions in the solar
 system neighborhood, at least at very high energy.  Still a galactic
 component may be important at intermediate energies (10--100 GeV)
 [5].  In addition, the discrete nature in space and time of the
 acceleration process can make unpredictable the spectrum above 100
 GeV and produce a positive curvature of the spectrum in log-log scale
 at intermediate energies [1].

\section{The direct measurements of CR electrons}

\begin{figure}
\includegraphics[width=0.5\textwidth]{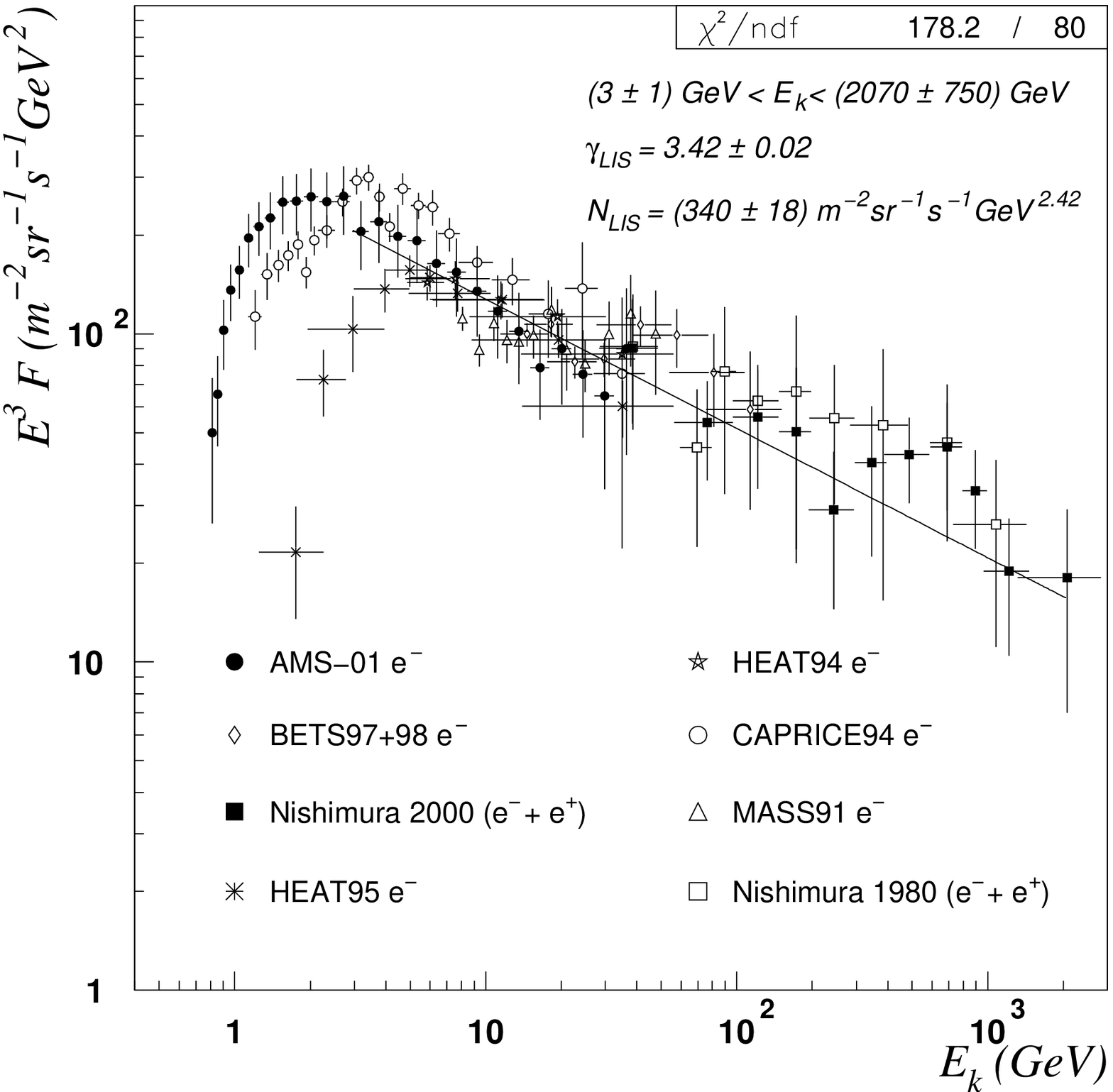} \hfill
\includegraphics[width=0.481\textwidth]{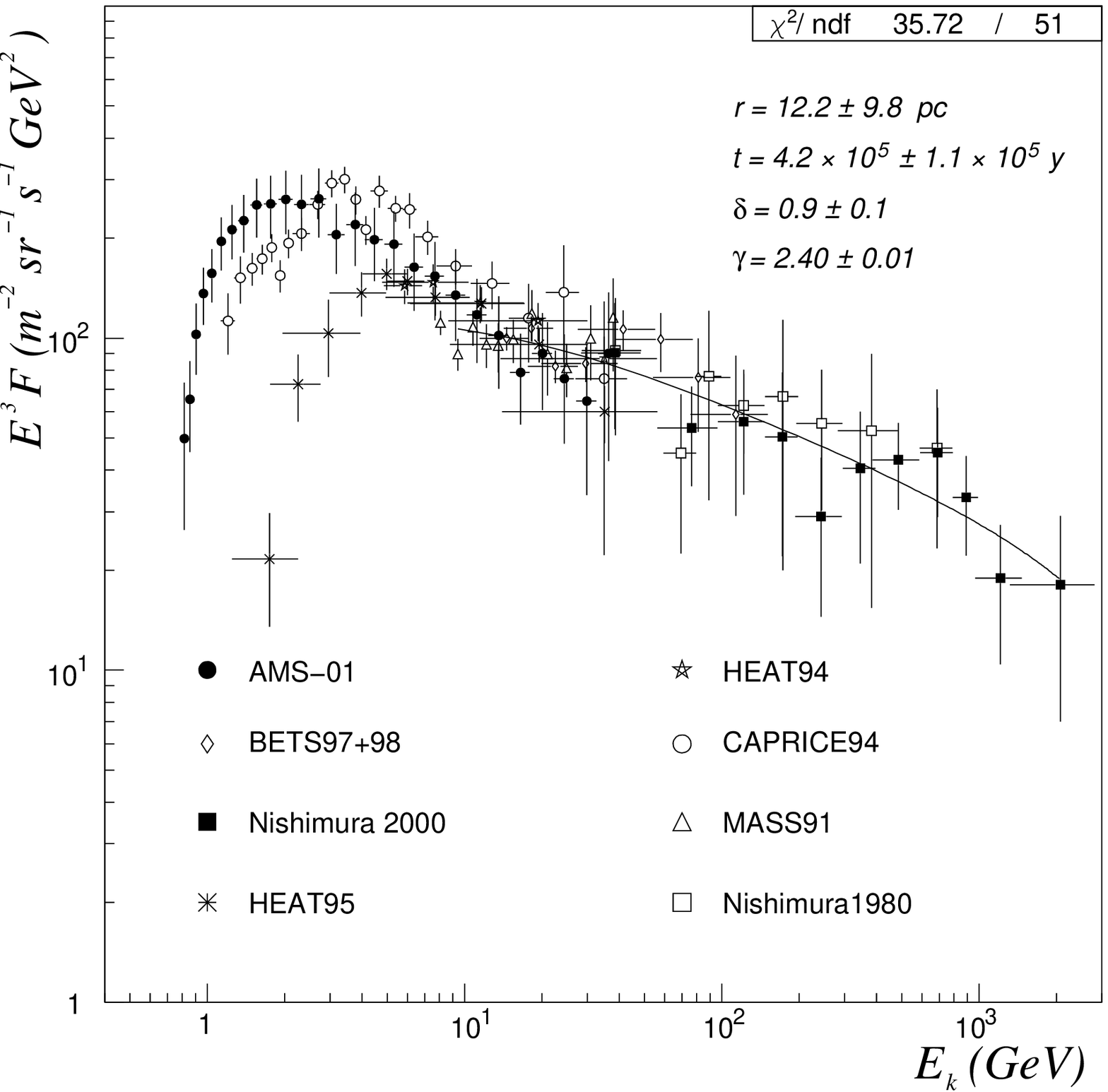}
\caption{Local interstellar spectrum of cosmic ray electrons with
  single power-law fit (left) and single source model fit (right).
  All data sets were re-normalized to the same value at 20
  GeV.}\label{lis}
\end{figure}

 The local interstellar spectrum (LIS) of CR electrons measured by
 recent experiments is shown in figure~\ref{lis}~ The correction for
 the solar modulation has been carried on with the spherical adiabatic
 model of Gleeson \& Axford [8].  In addition, all data points were
 re-normalized to the flux measured by AMS-01 at 20 GeV, in order to
 reduce the spread, assuming that systematic errors in the acceptance
 calculations do not affect the spectral index [9].

 The left panel shows that a a global fit with a single power-law from
 3 GeV up to 2 TeV seems to be not excluded even though the spread of
 the data set is still visible after the re-normalization.  The unique
 spectral index over this large energy range may be explained with the
 hypothesis that the measured spectrum is produced by a single nearby
 source, with spectral index 2.4.  In this case the shock compression
 ratio $R=\rho_2/\rho_1$, where $\rho_1$ and $\rho_2$ are the upstream
 and downstream densities respectively, can be estimated from the
 relation $\gamma \simeq (R+2)/(R-1)$ [10].  One obtains $R=3.14$, a
 value that is quite different from the number usually adopted in the
 models of CR acceleration.  The usual compression ratio $R=4.0$
 (obtained with high Alfv\'enic Mach numbers [10]) corresponds to a
 spectral index of the injection spectrum equal to 2.0, in agreement
 with the average spectral index ($\simeq 0.5$) of the diffuse
 synchrotron radiation in the Galaxy [1].  Howewer, a source spectral
 index of 2.4, corresponding to a synchrotron spectral index of 0.7,
 is stil compatible with some of the SNR of the Green's catalogue
 [11].

 The highest energy points may suggest a spectral change or a cut-off
 near 1 TeV, but the large error bars do not allow to draw any
 definite conclusion.  A possible fit of the measured spectra above 10
 GeV with the single source model of Atoyan et al.~[5] is shown in the
 right panel of figure~\ref{lis}~ The best fit is obtained with a
 recent and nearby source with age $(4.2 \pm 1.1) \times 10^5$ y and
 distance $12.2 \pm 9.8$ pc from the solar system, with injection
 spectral index 2.4.  Again, this result suggests that we observe
 cosmic ray electrons accelerated by a nearby single burst-like
 process, rather than multiple sources (as considered in ref.~[1]).
 In addition, this model suggests that the SN explosion happened not
 so long time ago, and that the value for $\delta$ is quite high
 compared to the usual value of 0.6 adopted in the literature, even if
 it is compatible with recent simulations by Maurin et al.~[7], who
 find a better agreement to the secondary/primary ratio if $\delta =$
 0.8--0.9.

 Kobayashi et al.~[4] took into consideration several nearby SNR,
 concluding that Vela (distance 250 pc, age 20 ky) should dominate the
 observed spectrum, considered a superposition of few single source
 models.  Howewer, Guetta \& Amato [3] considered the SNR whose TeV
 gamma rays has been detected and conclude that the high energy
 photons coming from Vela are likely due to hadronic processes instead
 of inverse Compton scattering of accelerated electrons with ambient
 photons.  In addition, our fit shows that the single source must be
 nearer than Vela: it is probably the last SN among those that
 contributed to the formation of the local bubble [12].  Thus the site
 where the measured CR electrons were accelerated is still a matter of
 debate.
\newline

\noindent
 The authors whish to thank L.~Foschini, S.~Cecchini, R.~Fanti and
 F.~Palmonari.

\section{References}

\re
1.  Pohl M. \& Esposito J.A.,
 ApJ 507 (1998) 327.
\re
2.  Reynolds S.P. \& Keohane J.W.,
 ApJ 525 (1999) 368;
    Hendrick S.P. \& Reynolds S.P.,
 ApJ 559 (2001) 903.
\re
3.  Guetta D. \& Amato E.,
 Astropart. Phys. 19 (2003) 403.
\re
4.  Kobayashi T. et al.,
 Adv. Space Res. 27, 4 (2001) 653.
\re
5.  Atoyan A.M. et al.,
 Phys. Rev. D52, 6 (1995) 3256.
\re
6.  Slane P. et al.,
 ApJ 548 (2001) 814;
    Dyer K.K. et al.,
 ApJ 551 (2001) 439;
    Kong A.K.-K. et al.,
 ApJ 580 (2002) L125.
\re
7.  Maurin D. et al.,b
 A\&A 394 (2002) 1039.
\re
8.  Gleeson L.J. \& Axford W.I.,
 ApJ 149 (1967) L115; ApJ 154 (1968) 1011.
\re
9.  Du Vernois M.A. et al.,
 ApJ 559 (2001) 296;
    M\"uller D.,
 Adv. Space Res. 27, 4 (2001) 659.
\re
10. Schlickeiser R.,
``Cosmic Ray Astrophysics'', Springer 2002.
\re
11. Green D.A.,
 `A Catalogue of Galactic Supernova Remnants (Dec.~2001 version)',
 Mullard Radio Astronomy Observatory, Cavendish Laboratory, Cambridge,
 United Kingdom (http://www.mrao.cam.ac.uk/surveys/snrs/).
\re
12. Ma\'{\i}z-Apell\'aniz J.,
 ApJ 560 (2001) L83;
    Ben\'{\i}tez N. et al.,
 Phys. Rev. Lett. 88, 8 (2002) 081101.

\endofpaper
\end{document}